# Thermal Analysis of Metal-Organic Precursors for Functional Cu:NiOx Hole Transporting Layer in Inverted Perovskite Solar Cells: Role of Solution Combustion Chemistry in Cu:NiOx Thin Films Processing

Apostolos Ioakeimidis [a], Ioannis T. Papadas [a,c], Eirini D. Koutsouroubi [b]  and Gerasimos S. Armatas [b] and Stelios A. Choulis [a,*]

[a] Molecular Electronics and Photonics Research Unit, Department of Mechanical Engineering and Materials Science and Engineering, Cyprus University of Technology, Limassol, Cyprus.
[b] Department of Materials Science and Technology, University of Crete, Heraklion 70013, Greece.
[c] Department of Public and Community Health, School of Public Health, University of West Attica, Athens, Greece.
[*] Correspondence: stelios.choulis@cut.ac.cy

**Abstract:** Low temperature solution combustion synthesis emerges as a facile method for synthesis of functional metal oxides thin films for electronic applications. We study the solution combustion synthesis process of Cu:NiO$_x$ using different molar ratios (w/o, 0.1 and 1.5) of fuel acetylacetone (Acac) to oxidizer (Cu, Ni Nitrates) as a function of thermal annealing temperatures 150, 200 and 300 °C. The solution combustion synthesis process, in both thin films and bulk Cu:NiO$_x$, is investigated. Thermal analysis studies using TGA and DTA reveal that the Cu:NiO$_x$ thin films show a more gradual mass loss while the bulk Cu:NiO$_x$ exhibits a distinct combustion process. The thin films can crystallize to Cu:NiO$_x$ at annealing temperature of 300 °C irrespective to the Acac/Oxidizer ratio whereas lower annealing temperatures (150 and 200 °C) produce amorphous materials. A detail characterization study of solution combustion synthesized Cu:NiOx including XPS, UV-Vis, AFM and Contact angle measurements is presented. Finally, 50 nm Cu:NiO$_x$ thin films are introduced as HTLs within the inverted perovskite solar cell device architecture. The Cu:NiO$_x$ HTL annealed at 150 and 200 °C provided PVSCs with limited functionality whereas efficient triple-cation Cs$_{0.04}$(MA$_{0.17}$FA$_{0.83}$)$_{0.96}$Pb(I$_{0.83}$Br$_{0.17}$)$_3$ based PVSCs achieved for Cu:NiO$_x$ HTLs annealed at temperature 300 °C.

## 1. Introduction

Perovskite solar cells (PVSCs) have witnessed significant progress related to power conversion efficiency within the last decade, climbing from 3.8% in 2009 to more than 24% in the year 2021 for single junction cells.[1,2] Some of the most promising hole transporting layers (HTLs) for inverted PVSCs are thin films of pristine or doped NiO$_x$ materials grown with various deposition methods. Thanks to the p-type semiconducting nature, high optical transmittance, enhanced electrical conductivity and deep-lying valence band (VB) that matches well with the VB of hybrid perovskite photoactive layer, NiO$_x$ emerges as an excellent hole transporting layer (HTL) material for inverted perovskite solar cells.[3–17]. Doping of NiOx with Copper (Cu) has been shown to improve the conductivity and charge collection properties resulting in higher PCE inverted perovskite solar cells. Jong H. Kim studied the performance of PEDOT:PSS, NiOx and 5% Cu doped NiOx (Cu:NiOx) HTLs relevant to inverted perovskite photovoltaic performance . Among them, Cu:NiOx exhibited the higher PCE (14.89 %) due to better valence band (VB) alignment with the perovskite active layer compared to PEDOT:PSS HTL, while in comparison to pristine NiOx the higher PCE was ascribed to increased Cu:NiOx electrical conductivity (8.4x10$^{-4}$ S.cm$^{-1}$) due to Cu doping.[18] Wei Chen et. al. has reported that doping of NiOx with 5% Cu induces a slight downshift of VB from 5.16 to 5.25 eV and at the same time increases both carrier concentration and hole mobility of the Cu:NiOx HTL. The Cu:NiOx HTL based inverted perovskite solar cells provided an increased PCE of 18.01% compared to pristine NiOx HTL based solar cells (16.68 %) retaining 95% of the PCE after 1000 h storage in air. [19] However, NiO$_x$ HTL derivatives prepared by sol-gel method usually need to be annealed at relative high temperatures (over 400 °C) in order to achieve high crystallinity.[20–23] Sol-gel reactions are endothermic, and thus require high external thermal energy to form metal oxide lattices and remove organic residuals.[24–27] This high processing temperature increases the cost of device fabrication and also inhibits the implementation of printing manufacturing for producing next generation solution processed photovoltaics on conventional flexible substrates.

Combustion synthesis methods have been reported by many researchers as a beneficial route of synthesis of high crystalline metal oxides, at lower temperatures than those used for sol-gel reactions. The combustion process exothermic reaction provides a lower transition energy to form the metal oxide crystal lattices, avoiding the need for high thermal energy. Thus, the solution combustion synthesis (SCS) method is being adopted a for the synthesis of metal oxides electronic thin films due to its cost-effectiveness, simplicity, enhanced electronic material functionality and relatively lower required processing annealing temperatures.[28–36]

Over the last years, there have been many reports for the use of solution combustion synthesis of metal oxide such as the Cu:CrO$_x$ and ZnO for charge transporting layers in perovskite solar cells and Amorphous Indium Gallium Zinc oxide (IGZO) for metal oxide thin film transistors applications . [29,31,37,38] We have reported the solution combustions synthesis of pristine and co-doped (Cu,Li) NiCo$_2$O$_4$ films applying a 300 °C annealing temperature and incorporating them as high performance HTLs in MAPbI$_3$ based inverted perovskite solar cells.[39,40] We have also shown the improvements of PVSC's thermal stability based on SCS Cu:NiOx HTL by treatment of Cu:NiOx with β-alanine showing a T80 of 1000 h under heat conditions (60 °C, N$_2$), as well as the improvement of humidity degradation resistance for SCS NiO$_x$ based PVSC with the addition of 1% Nitrobenzene within the perovskite active layer .[41,42] Jae Woong Jung et. al. has reported the solution combustion synthesis of the Cu:NiOx film at 150 °C and implemented as HTL in MAPbI$_3$ perovskite solar cell. They showed that the combustion synthesized Cu:NiOx resulted in better power conversion efficiency (PCE) perovskite solar cells compare to devices containing a typical sol-gel synthesized Cu:NiO$_x$ HTL.[43] Other reports on solution combustion synthesis of pristine and doped NiO$_x$ have applied a range of annealing temperatures for the fabrication of HTLs for efficient perovskite solar cells.[44–49] For example, Ziye Liu et. al. reported the fabrication of MA$_{1-y}$FA$_y$PbI$_{3-x}$Cl$_x$ perovskite solar cell using solution combustion synthesized NiOx achieving a high PCE of more than 20 %. The applied temperature of NiOx solution combustion synthesis for high efficiency devices was 250 °C while for 150 °C annealing temperature the devices exhibited a rapid deterioration of their PCE.[47] Ao Liu et. al., demonstrated the solution combustion synthesis fabrication of optimized 5% Cu doped NiO$_x$ films for use in TFT applications exhibiting excellent electrical performance.[50] Yi-Huan Li et. al. applied 300 °C annealing temperature for the solution combustion synthesis of Cu:NiOx showing that it can be used as efficient hole injection layer for the fabrication of high performing OLEDs.[51] Thus, solution combustion synthesis metal oxide thin films have emerged as a facile method for fabrication of functional metal-oxide based charge selective contacts for electronic applications. Nevertheless, there have been reports that the combustion synthesis of thin electronic films differ from the corresponding bulk analogues, and thus low annealing temperature combustion synthesis of thin films cannot be deduced by the bulk material behavior, even suggesting that low temperature combustion synthesis is unlikely to occur during the processing of thin film precursors.[52–57] Thus, the processing annealing conditions for the solution combustion synthesis of functional Cu:NiOx HTLs needs further investigation.

The aim of this paper is to identify the Cu:NiOx HTL annealing processing conditions and to examine the fuel to oxidizer ratio for efficient inverted PVSCs. We study in detail the Cu:NiO$_x$ thin films solution combustion synthesis process as a common metal oxides HTL that is widely used in inverted perovskite solar cells. Specifically, the reported results investigate the effect of the thermal annealing temperature (150, 200 and 300 °C) as well as the fuel [acetyl acetonate (Acac)] to oxidizer (Cu and Ni nitrates) ratio [without (w/o), 0.1 and 1.5] in the combustion synthesis process of Cu:NiO$_x$. The study is performed at Cu:NiOx films, with various final? thickness (50, 200, 300 nm) and for drop-casted bulk Cu:NiOx analogues (with thickness in the range of a few microns). The crystal growth process is studied by performing thermogravimetric analysis (TGA) and the crystallinity of the corresponding Cu:NiOx materials is examined by X-ray diffraction (XRD). Furthermore, we characterized the thin Cu:NiOx films properties processed under the above-mentioned conditions using, XPS, contact angle, AFM and UV-Vis spectroscopy techniques. Finally, we evaluated the impact of SCS based Cu:NiO$_x$ thin film properties on the PCE performance of inverted PVSCs containing Cu:NiO$_x$ HTL and the triple cation perovskite [Cs$_{0.04}$(MA$_{0.17}$FA$_{0.83}$)$_{0.96}$ Pb(I$_{0.83}$Br$_{0.17}$)$_3$] as photoactive layer.

## 2. Materials and Methods

*Materials:* Pre-patterned glass-ITO substrates (sheet resistance 4Ω/sq) were purchased from Psiotec Ltd. All the other chemicals used in this study were purchased from Sigma Aldrich.

*Cu:NiO$_x$ solution combustion synthesis:* In a typical synthesis, 0.95 mmol Ni(NO$_3$)$_2$.6H$_2$O and 0.05 mmol Cu(NO$_3$)$_2$.3H$_2$O were dissolved in 10 ml 2-methoxyethanol with different concentrations of fuel acetylacetonate to the solution and the mixture was further stirred for 1 h at room temperature. Then, the samples were dried at 80 °C for 5 min and annealed at 150 °C, 200 °C or 300 °C in air for 1 hour. The chemical reaction formula of the solution combustion synthesis is the above :

$Ni(NO_3)_2 \cdot 6H_2O + Cu(NO_3)_2 \cdot 3H_2O + C_5H_8O_2 \Rightarrow Cu:NiO_x (s) + \uparrow H_2O + \uparrow CO_2 + \uparrow N_2$

*Samples preparations for TGA, AFM, UV-Vis, contact angle analysis:* For the thermogravimetric analysis (TGA) of combustion synthesis behavior of the Cu:NiO$_x$ films the samples were fabricated onto Alumina disk substrates by blade coating in air. For the AFM, UV-Vis and contact angle measurements the Cu:NiO$_x$ films were fabricated by doctor blade on quartz substrates.

*Device fabrication:* ITO-patterned glass substrates were cleaned using an ultrasonic bath for 10 min in acetone followed by 10 min in isopropanol. The Cu:NiO$_x$ precursors were prepared, and blade coated on ITO substrates as described in Cu:NiOx solution combustion synthesis section. For the preparations of the triple cation Cs$_{0.04}$(MA$_{0.17}$FA$_{0.83}$)$_{0.96}$ Pb(I$_{0.83}$Br$_{0.17}$)$_3$ perovskite solutions a previous reported method was used.[58] The perovskite films were fabricated on top of Cu:NiO$_x$ inside a N$_2$ atmosphere glovebox by spin-coating at 5000 rpm for 35 s and after 10 sec. 300 mL of ethyl acetate were dropped onto the spinning substrate as the anti-solvent to achieve the rapid crystallization of the films. The resulting perovskite films were annealed at 100 °C for 60 min. As electron transporting layers a PC$_{60}$BM film was coated on top of perovskite inside the glovebox using spin coating at 1000 rpm for 30 s from a 20 mg/ml chlorobenzene solution. To complete the devices, 7 nm of bathocuproine (BCP) were thermally evaporated followed by 80 nm of Ag. The schematic illustration of SCS based Cu:NiO$_x$ HTL and PVSC fabrication process and the corresponding device structure are presented in Figure 1.

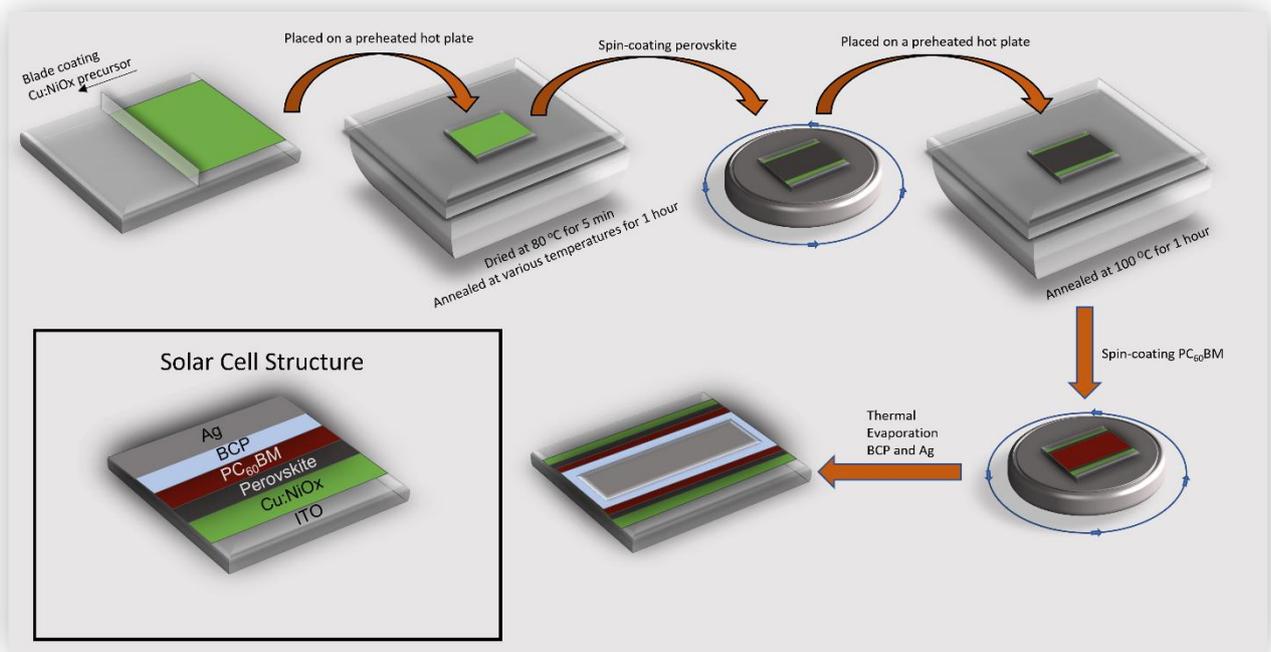

**Figure 1.** Schematic illustration of SCS based Cu:NiOx HTL and perovskite device fabrication process and the layer structuring.

*Characterization:* Thermogravimetric analysis (TGA) and differential thermal analysis (DTA) was performed on a Shimadzu. Sample were heated up to 400 °C in air atmosphere (200 mL min$^{-1}$ flow rate) with a heating rate of 10 °C min$^{-1}$ and as reference material analysis was used alumina (Al$_2$O$_3$), a substance with the same thermal mass as the sample. Powder X-ray diffraction (XRD) patterns were recorded on a PANalytical X´Pert Pro X-ray diffractometer with a Ni-filtered Cu K$\alpha$ source ($\lambda$ = 1.5418 Å), operating at 45 kV and 40 mA. X-ray photoelectron spectroscopy (XPS) analysis conducted on a SPECS spectrometer using a Phoibos 100 1D-DLD electron analyzer and an Al K$\alpha$ radiation as the energy source (1486.6 eV). Binding energy values were corrected for charging by assigning a bending energy of 284.8 eV to the C 1s signal of adventitious carbon. For UV-Vis absorption and atomic force microscopy (AFM) measurements the films were fabricated on quartz substrates. UV-Vis absorption measurements were performed with a Schimadzu UV-2700 UV-Vis spectrophotometer. AFM images were obtained using a Nanosurf easy scan 2 controller applying tapping mode. The thickness of the films was measured with a Veeco Dektak 150 profilometer. Contact angle (CA) measurements were performed using a KRUSS DSA 100E drop analysis system. The current density-voltage (J/V) characteristics were characterized with a Botest LIV Functionality Test System. For illumination, a calibrated Newport Solar simulator equipped with a Xe lamp was used, providing an AM1.5G spectrum at 100 mW/cm$^2$ as measured by a certified oriel 91150V calibration cell. A shadow mask was attached to each device prior to measurements to accurately define 0.09 cm$^2$ device area.

## 3. Results

### 3.1. TGA results of Cu:NiO$_x$ (films versus bulk precursors)

The synthesis behavior of the Cu:NiO$_x$ thin films and corresponding bulk mixtures onto alumina disc substrates was examined through thermogravimetric analysis (TGA) and differential thermal analysis (DTA). Figures 2(a) and 2(b) present the TGA and the corresponding DTA curves of the 50 nm thick Cu:NiO$_x$ films prepared with different molar ratio of fuel (Acac) to oxidizer (Cu and Ni nitrates), namely without (w/o) Acac, 0.1 and 1.5, and drying the films at 80 °C. The Cu:NiO$_x$ film w/o Acac shows a mass loss near ~130 °C, indicating the thermal instability of this precursor in the absence of any fuel additive. For the Cu:NiO$_x$ films prepared with 0.1 and 1.5 Acac/oxidizer molar ratio, the TGA profiles show similar thermal decomposition behavior, exhibiting a gradual mass loss after T > ignition temperature ($T_{ig}$). This is not consistent with a combustion process that occurs to the combustible precursors as will be shown below. Near 300 °C, the TGA profiles for all samples display an intense mass loss, which is associated with an exothermic peak on the DTA curve. This is attributed to the decomposition of metal complexes and the crystallization of the Cu:NiO$_x$ oxide. Thus, films present two stages of mass loss, at ~130 °C and ~300 °C for drying at 80 °C irrespective of Acac addition in precursor solution. The corresponding DTA results are presented in Figure 2(b). For the sample w/o Acac an negligible broad exothermic peak is observed at ~130 °C and another broad exotherm peak around ~300 °C. For the samples containing 0.1 and 1.5 Acac low intensity abrupt exothermic peaks are observed at ~130 °C and broad exothermic peaks around ~300 °C. Thus, the addition of Acac in the precursor induce a limited combustion process by reacting with a part of the oxidizer at ~130 °C facilitating the removal of the organic residuals.

The mass loss at ~130 °C of the sample w/o Acac suggests that the 2-methoxy ethanol, except its role as solvent, could also behave as a fuel. To support this, higher drying temperature (100 °C) for 1 hour was applied to the film's synthesis where most of 2-methoxy ethanol was evaporated. In this case, the TGA curve showed no mass loss and the corresponding DTA curve exhibited analogous behavior without any endothermic or exothermic reaction at 130 °C, while for the film prepared with a Acac/oxidizer molar ratio of 0.1 a marginal mass loss occurs during the first stage of combustion (at ~130 °C). This means for the films that non-combustive Cu:NiOx precursors require high temperatures of over 300 °C for the complete conversion of the precursors into the metal oxide lattice.

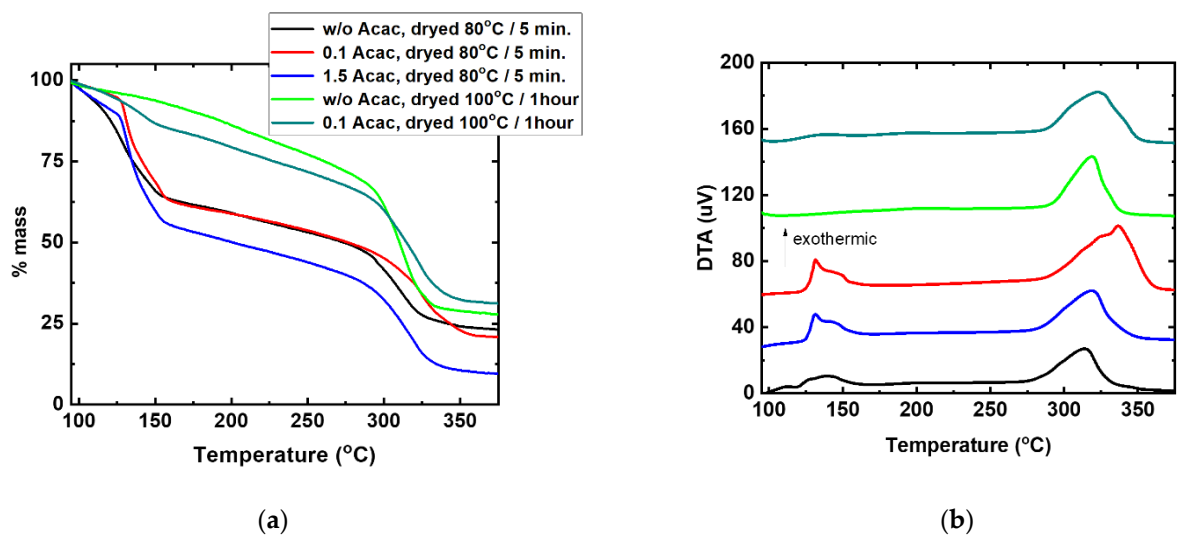

(a)  (b)

**Figure 2.** (a) TGA curves of precursor films without (w/o) fuel and containing 0.1 and 1.5 molar ratio of fuel (Acac) to oxidizer (Cu, Ni nitrates) in 2-methoxy ethanol dried at 80 and 100 °C, and (b) the respective DTA curves.

Further, we compared the combustion synthesis behavior of Cu:NiO$_x$ thin films (50, 200, 300 nm) and bulk analogues (thickness range of a few microns). TGA profiles (Figure 3(a)) show that by increasing the thickness of the film, a more intense gradual mass loss occurs at ~130 °C, with the second mass loss at ~300 °C becoming less prominent, while comparatively full combustion could occur at bulk materials at ~130 °C. Thus, we can infer that the mass of the precursor has a significant impact on the complete combustion synthesis reactions. The DTA results, in Figure 3(c), are in accordance with TGA profile where the raise of thickness in Cu:NiOx precursors layers shows a more intense exothermic peak at 130 °C; this corresponds to almost complete mass loss, decreasing gradually the exothermic peak at around 300 °C. Additionally, the absence of fuel (Acac) and the impact of solvent was examined once again during the combustion process for bulk materials. As observed in TGA profile (Figure 3(b)), a rapid mass loss occurs at ~130 °C w/o Acac for samples dried at 80 and 100 °C for 5 min, respectively. The corresponding DTA curves (Figure 3(d)) show single sharp exotherms at ~ 130 °C that corresponds exactly to the abrupt mass loss in the TGA (Fig. 3b); this process is ample to lead the reaction rapidly to completion for metallic Ni formation (Fig. 5). In contrast, in preheated sample at 100 °C for an extended period (48 h), where most of the solvent was evaporated, combustion reaction could not occur. This sample exhibits only an intense exothermic peak around ~300 °C, that corresponds to the crystal phase formation of NiO. These observations are in agreement with previous reports that the organic solvent 2-methoxy ethanol plays a dual role of acting both as a solvent and also as a fuel in addition to Acac for the formation of the metal oxide lattices by the solution combustion synthesis.[59]

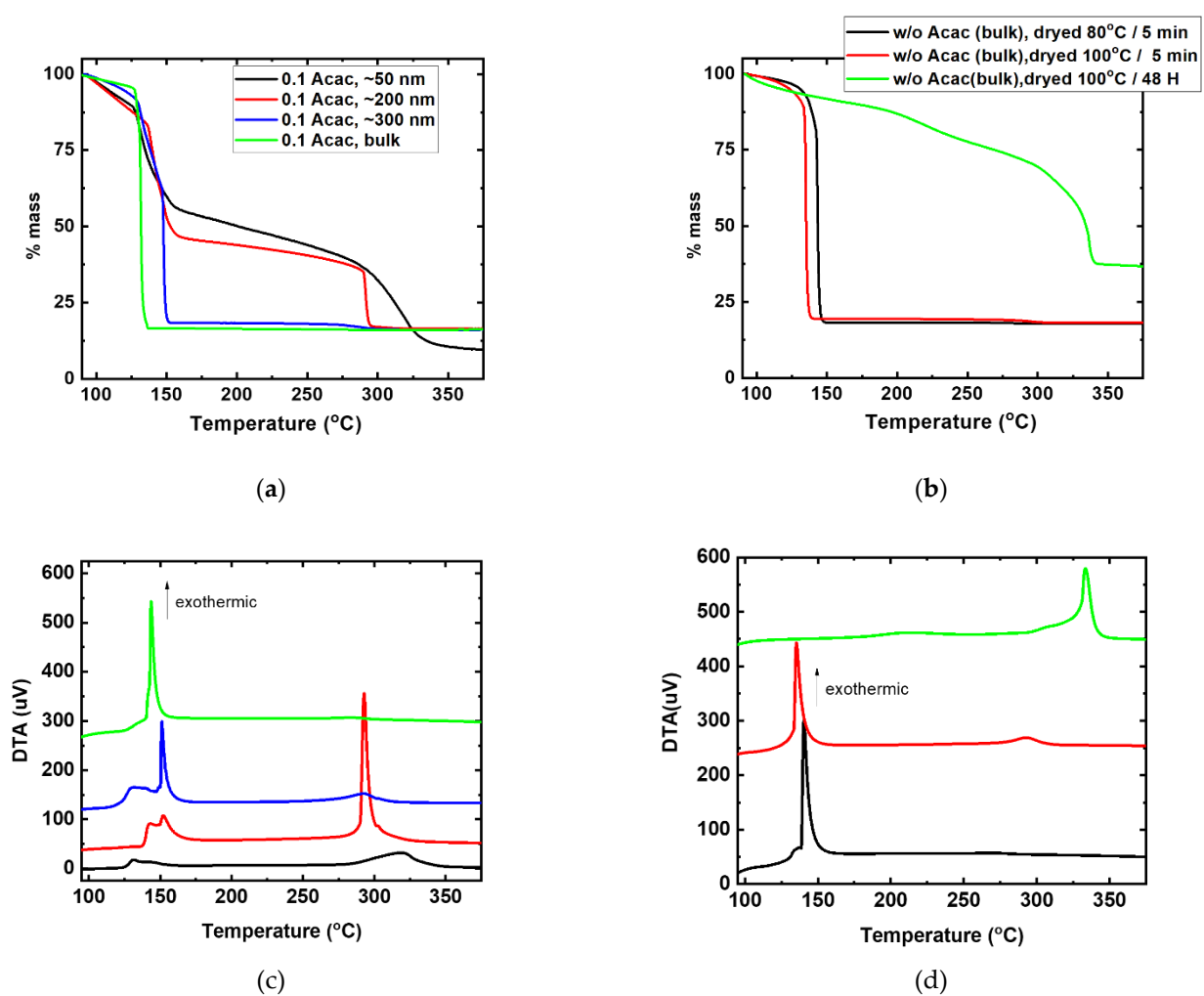

**Figure 3.** (a) TGA curves of different thickness (50, 200, 300 nm and bulk) films containing 0.1 molar ratio of fuel (Acac) to oxidizer (Cu, Ni nitrates) in 2-methoxy ethanol and (b) TGA curves of combustion-synthesized bulk samples prepared from precursor with Cu and Ni nitrates but without (w/o) Acac and 2-methoxy ethanol as solvent dried at 80 °C and 100 °C for 5 min and at 100 °C for 48 h. The respective DTA curves for (c) different thickness (50, 200, 300 nm and bulk) and (d) bulk samples.

*3.2. XRD results of Cu:NiOx films and bulk precursors*

The crystallinity of the SCS Cu:NiOx thin films (identical to the Cu:NiOx HTLs that were used within the inverted PVSCs) was examined using X-ray diffraction (XRD) analysis. Figure 4(a-c) presents the XRD patterns of the films prepared using w/o, 0.1 and 1.5 Acac and annealing temperatures of 150, 200 and 300 °C. The crystal phase of NiO can be obtained for annealing temperature 300 °C regardless of the containing amount of Acac, while for 150 and 200 °C no crystal phase was detected. For 300 °C annealing temperature, the characteristic diffraction peaks of NiO appeared at 2θ = 37.20°, 43.0°, 62.87° and 75.20°, which can be indexed to the cubic crystal structure of NiO as (111), (200), (220) and (311) planes, respectively (JCPDS No. 01-089-5881). For the film containing 1.5 Acac and annealed at 300 °C, the formation of mixed crystal phases of NiO and metallic Ni was observed. Specifically, XRD patterns, along with the NiO diffractions, reveal additional peaks at 2θ = 44.0°, 52.3° and 76.5° assigned to (111), (200) and (220) planes, respectively, of the face-centered cubic (FCC) phase of Ni (JCPDS No.87-0712).

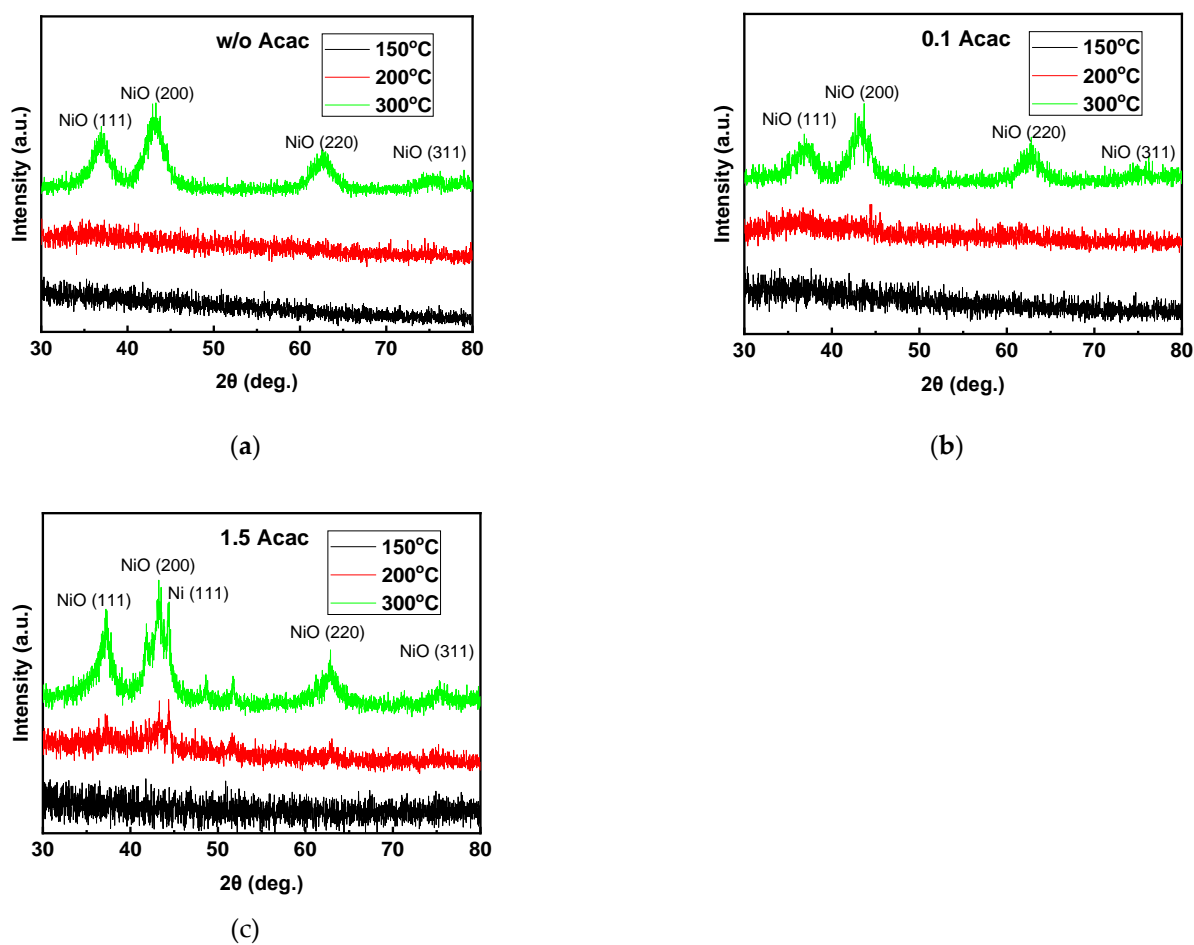

**Figure 4.** XRD patterns of the combustion synthesis of precursor films containing (a) w/o fuel (Acac), (b) 0.1 and (c) 1.5 molar ratio of fuel (Acac) to oxidizer (Cu, Ni nitrates) annealed at 150, 200 and 300 °C.

The crystallinity of as-prepared materials obtained by combustion reaction of the bulk precursors was also examined using XRD analysis (Figure 5). Specifically, different initial molar ratios of the fuel to oxidizer (w/o, 0.1 and 1.5 Acac) at 200 °C annealing temperature of the bulk precursors were compared. The XRD results show significant improvement in the crystallinity of the bulk Cu:NiO$_x$ compared to the corresponding thin films. Moreover, the required annealing temperature for the crystal phase formation is significantly reduced (as was also evidenced by TGA analysis) when using solution combustion synthesis of bulk materials as compared to the thin films. In the case of samples synthesized by bulk precursors, mixed crystal phases of metallic Ni (dominant species) and metal oxide NiO (residual species) were obtained, regardless of the molar ratio of fuel to oxidizer precursors. Specifically, XRD patterns reveal intense peaks appeared at 44.0°, 52.3° and 76.5° assigned to the (111), (200) and (220) planes, respectively, of face-centered cubic Ni (JCPDS No. 87-0712) (main product). The presence of Ni phase is an indication of combustion with a rich fuel precursor, so even in the precursor w/o Acac the Ni phase is the main product implying again that 2-methoxy ethanol plays a dual role of acting both as a solvent and fuel.[27] The high crystallinity of bulk materials, as evidenced by the sharper diffractions in XRD patterns, is attributed to the release of high energy in the exothermic reaction that occurred by solution combustion synthesis (SCS). Thus, in agreement with the findings from TGA measurements, the complete combustion occurs mainly due to the bulk precursor material, which has a higher mass compared to corresponding films.[56]

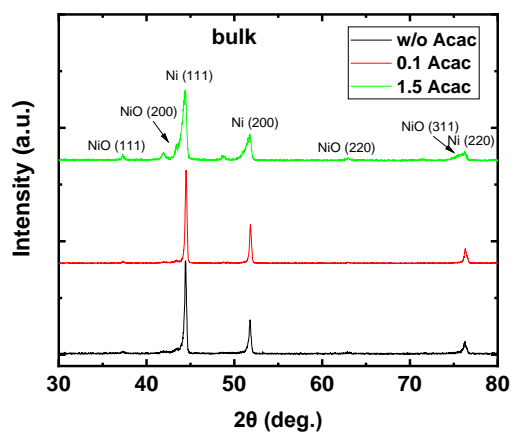

**Figure 5.** XRD patterns of combustion-synthesized samples (bulk) prepared from precursors containing w/o, 0.1 and 1.5 molar ratio of fuel (Acac) to oxidizer (Cu, Ni nitrates) annealed at 200 °C.

*3.3. Cu:NiOx thin films characterization*

X-ray photoelectron spectroscopy (XPS) was employed to investigate the chemical state of the Cu:NiO$_x$ surface. The XPS survey scans of the Cu:NiOx films synthesized with 0.1 Acac and annealed at 200 and 300 °C evidenced the presence of Ni, Cu, O and C elements (Figure S1). In the Cu:NiOx film annealed at 200 °C, the N1s spectrum indicated the presence of some reduced nitrogen (399.8 eV ), and NO$_2^-$ (403.6 eV) and NO$_3^-$ (406.8 eV) containing species (Figure 6a), while the N1s scan of the 300 °C annealed film showed the exist of reduced nitrogen (399.0) and NO$_x$ (405.6 eV) residues (Figure 6b).[60] For the film annealed at 200 °C, the spectrum of the Ni 2p region (Figure 6c) showed a double peak at 856.3 eV (Ni 2p$_{3/2}$) and 874.0 eV (Ni 2p$_{1/2}$) binding energies, accompanying with shake-up satellite peaks at 862.0 eV and 879.7 eV, which are characteristic of Ni$^{2+}$–oxygen bonded complexes, possibly in the form of Ni(acac)$_2$.[61] While the XPS Ni 2p spectrum of the 300 °C annealed film (Figure 6d) indicated the presence of NiO, showing a characteristic double peak at 854.6 eV (Ni 2p$_{3/2}$) and 872.3 eV (Ni 2p$_{1/2}$) binding energies (spin-orbit splitting of 17.7 eV) along with shake-up satellite peaks at 861.2 and 878.8 eV.[62] Furthermore, the XPS Cu 2p spectrum of the 300 °C annealed film (Figure 6e) exhibited a double peak at 934.2 eV and 953.9 eV due to the Cu 2p$_{3/2}$ and Cu 2p$_{1/2}$ core level components of the CuO (Cu 2p$_{3/2}$: 934.7 eV and Cu 2p$_{3/2}$: 954.5 eV for the 200 °C annealed film, Figure 6f), consistent with other reports.[63,64] As for the broad signals located at 940.6 and 953.9 eV (943.0 eV for the 200 °C annealed film), they are assigned to the shake-up satellite peaks of paramagnetic Cu$^{2+}$. The incorporation of Cu$^{2+}$ ions into the NiO lattice was also verified by the Auger $\alpha$ parameter, that is, the kinetic energy of the Cu L$_3$M$_{4,5}$M$_{4,5}$ Auger peak plus binding energy of the Cu 2p$_{3/2}$ peak. For the 300 °C annealed film, the Auger parameter is calculated to be 1851.8 eV, which respects the existing phase of CuO.[65] Quantitative analysis from the XPS spectra also indicated that the film annealed at 200 °C contains 6.51 wt.% CH$_x$O$_y$ and 3.32 wt.% NO$_x$ containing organic compounds, while the corresponding remnants for the 300 °C annealed film was found to be 0.64 wt.% and 0.33%, respectively, see Table S1. The higher amount of remnants found in 200 °C annealed film suggests the incomplete combustion reaction of Cu:NiOx precursors, in agreement with the TGA results (Figure 2). Also, the Cu atomic concentration (Cu doping level) in the Cu:NiOx films annealed at 200 and 300 °C was found to be 5.64 % and 5.84 %, respectively, very close to the nominal composition.

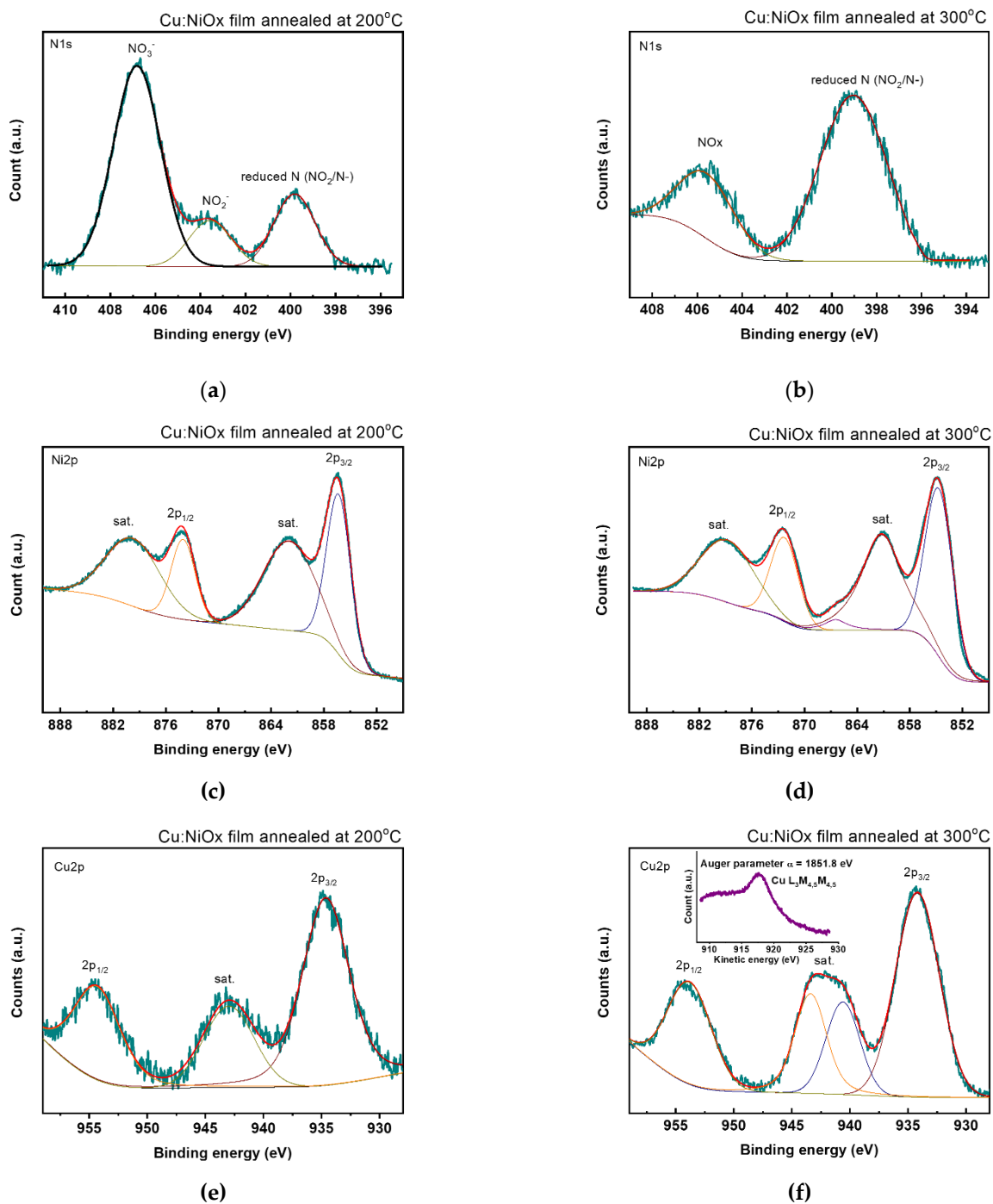

**Figure 6.** XPS spectra of the (a) ,(b) N 1s , (c),(d) Ni 2p and (e),(f) Cu 2p region of the Cu:NiOx films fabricated from precursor containing 0.1 Acac and annealed at 200ºC and 300 ºC. Inset of panel (f): the Cu L$_3$M$_{4,5}$M$_{4,5}$ Auger XPS spectrum

Furthermore, we examined the film topography of Cu:NiO$_x$ films fabricated on quartz substrates by contact angle and UV-Vis spectroscopy. Figure 7 shows the film morphology using AFM for the films synthesized from precursors containing 0.1 Acac and annealed at 150 and 300 ºC, respectively. It is clearly observed that the film treated at 150 ºC shows a featured structure of large particles, due to the presence of residues, with a surface roughness of 1.5 nm (Figure 7a). On the other hand, the film treated at 300 ºC does not show structured features due to the small size of the Cu:NiO$_x$ particles, exhibiting a surface roughness of 0.7 nm (Figure 7b). The final thickness of the film annealed at 150 and 300 ºC is ~80 and ~50 nm, respectively, due to considerable amount of residues remained into 150 ºC annealed film, as shown above through XPS analysis for films synthesized at low annealing temperatures.

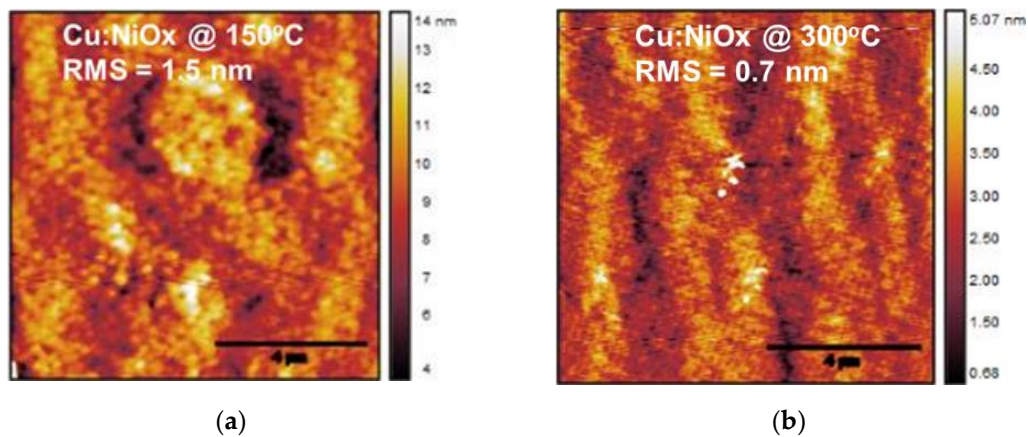

(**a**)                (**b**)

**Figure 7**. AFM images of Cu:NiO$_x$ films fabricated on quartz substrates from precursor containing 0.1 molar ratio of fuel (Acac) to oxidizer (Cu, Ni nitrates) annealed at **(a)** 150 and **(b)** 300 ºC.

The UV-vis absorption spectrum (Figure 8) of the films treated at 150 and 200 ºC show no prominent absorption due to the amorphous nature of metal oxides, while for 300 ºC annealing film the absorption onset at ~400 nm (~3.1eV) and strong absorption at ~350 nm (~3.5 eV) is ascribed to the crystalline Cu:NiO$_x$ phase.

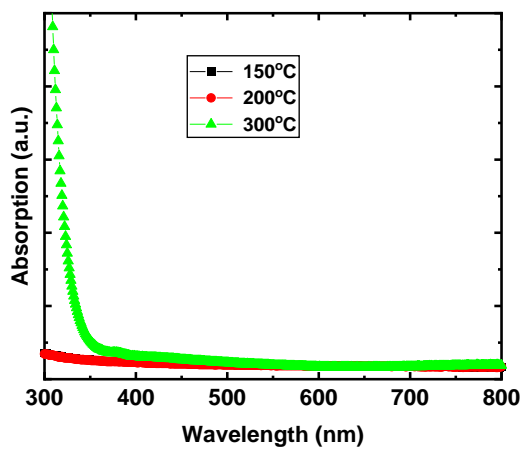

**Figure 8.** UV-Vis absorption of Cu:NiO$_x$ films fabricated on quartz substrates from precursor containing 0.1 molar ratio of fuel (Acac) to oxidizer (Cu, Ni nitrates) annealed at 150, 200 and 300 ºC.

The contact angle of water was measured on films annealed at 150, 170 and 200 ºC, using fuel to oxidizer ratio 0 (w/o), 0.1 and 1.5 (see Supporting Figure S2), and the measured values are plotted in Figure 9. All the contact angles are higher than 60 deg. irrespective to fuel concentration, in contrast to the contact angle of Cu:NiO$_x$ film (0.1 molar ratio of Acac to oxidizer) annealed at 300 ºC which is substantially lower (20 deg.), see Supporting Figure S3. Thus, we infer that the remnants (see XPS analysis) in the low temperature treated film form a Cu:NiO$_x$ surface with moderate wettability, while the films annealed at 300 ºC where the surface is almost free from remnants show an improved wettability.

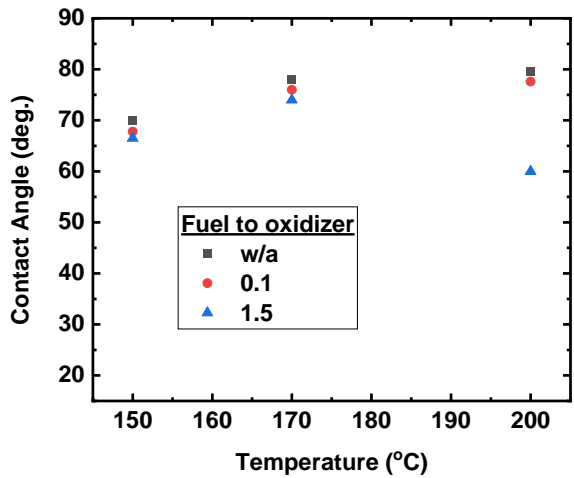

**Figure 9.** A graph presenting the contact angles of water on films prepared using precursor containing w/o Acac, 0.1 and 1.5 molar ratio of fuel (Acac) to oxidizer (Cu, Ni nitrates) annealed at 150, 170 and 200 °C.

*3.4. J-V characterization of Cu:NiO$_x$ films as HTLs in planar p-i-n PVSCs*

To evaluate the functionality of the different Cu:NiO$_x$ films as HTLs in solar cells, 50 nm thick Cu:NiO$_x$ films, synthesized using the previous described conditions, were implemented in inverted perovskite solar cells with structure ITO/Cu:NiO$_x$/perovskite/PC$_{60}$BM/BCP/Ag and the J-V device characteristics under 1 sun simulated light were recorded.

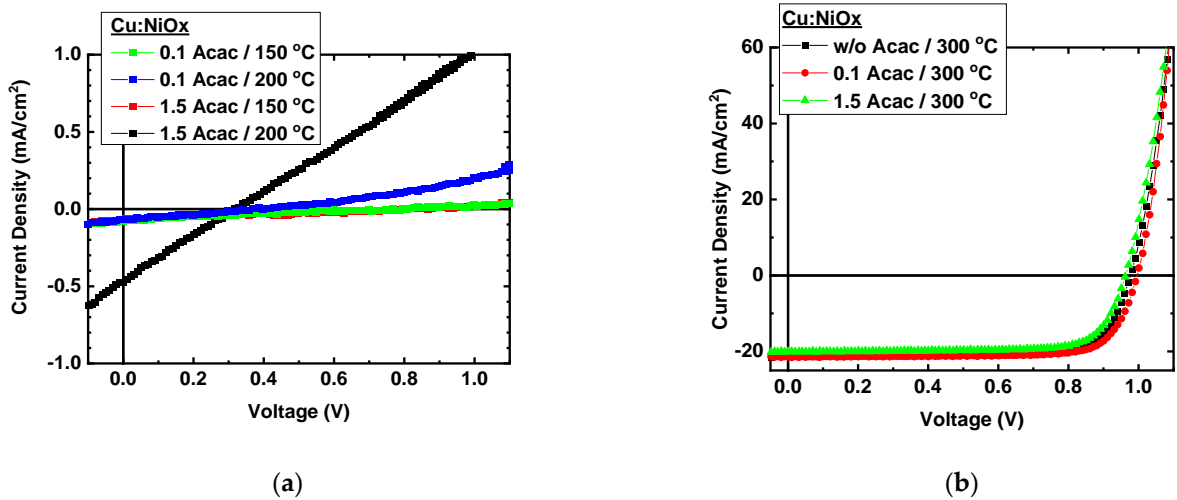

(**a**)                                           (**b**)

**Figure 10.** J-V curves of ITO/Cu:NiO$_x$/perovskite/PC$_{60}$BM/BCP/Ag devices under 1 sun simulated light for Cu:NiO$_x$ films fabricated from precursor containing (a) 0.1 and 1.5 molar ratio of fuel (Acac) to oxidizer (Cu, Ni nitrates) annealed at 150 and 200 °C, and from precursor containing (b) w/o Acac, 0.1 and 1.5 molar ratio of fuel (Acac) to oxidizer (Cu, Ni nitrates) annealed at 300 °C.

As it is presented in J-V curves of Figure 10(a) the devices that incorporate Cu:NiO$_x$ films annealed at 150 °C and 200 °C exhibited a very limited functionality. All the devices show low V$_{oc}$ in the range of 0.3 V and the generated current is below 1 mA/cm². The device with 1.5 molar ratio Acac to oxidizer and annealed at 200 °C shows almost linear response of the current density to the sweeping voltage, which can be attributed to partially formed metallic Ni, as can be inferred by the corresponding XRD results in Figure 4(c). In contrast, the inverted perovskite solar cells which incorporate Cu:NiO$_x$ HTLs prepared from precursor solutions containing w/o, 0.1 and 1.5 Acac annealed at 300 °C delivered higher efficiency. In Figure 10(b), the J-V curves of the best performing devices under 1 sun simulated light are illustrated and the extracted solar cell parameters of the studied devices are presented in Table 2 – in brackets are the average values of 12 devices for each batch. Regarding the impact of fuel to oxidizer ratio on the devices PCE, the devices with Cu:NiO$_x$ HTL where the precursor contained no fuel (w/o Acac) and 0.1 Acac ratio showed similar PCE values and the devices which incorporate Cu:NiO$_x$ film synthesized with 1.5 Acac/oxidizer ratio shows a reduced V$_{oc}$ and J$_{sc}$ efficiency, resulting in lower PCE.

**Table 2.** Extracted solar cell parameters of the best ITO/Cu:NiO$_x$/perovskite/PC$_{60}$BM/BCP/Ag devices. In bracket the average values of 12 devices for each batch are shown.

| Sample | $V_{oc}$ (V) | $J_{sc}$ (mA/cm$^2$) | FF (%) | PCE (%) |
|---|---|---|---|---|
| w/o Acac | 0.98 (0.97) | 21.11 (20.64) | 77.1 (72.3) | 15.97 (14.48) |
| 0.1 Acac | 0.99 (0.97) | 21.40 (20.75) | 78.2 (73.8) | 16.58 (14.85) |
| 1.5 Acac | 0.96 (0.94) | 20.03 (19.64) | 77.3 (73.1) | 14.90 (13.50) |

The experimental results presented within this manuscript using a triple cation Cs$_{0.04}$(MA$_{0.17}$FA$_{0.83}$)$_{0.96}$Pb(I$_{0.83}$Br$_{0.17}$)$_3$ perovskite formulation infer that the Cu:NiO$_x$ oxide's precursor films that were annealed at temperatures of 150 and 200 °C produce electronic films that cannot function as HTLs for efficient inverted perovskite solar cells. This is ascribed to the incomplete combustion that results in amorphous Cu:NiO$_x$ films with remnants. This result is in agreement with previous report where amorphous NiO$_x$ showed limited functionality as HTL when applied in organic solar cells.[66] On the other hand, as shown above pure crystalline phase of Cu:NiO$_x$ HTL obtained with annealing at 300 °C for the precursors w/o and 0.1 ratio of Acac/oxidizer whereas for 1.5 ratio metallic Ni are likely to be present within Cu:NiO$_x$ films (as indicated within the XRD pattern in Figure 4c). The pure crystalline phases (w/a and 0.1 ratio Acac/oxidizer) of Cu:NiOx resulted to better PCEs 15.97 % (average 14.48 %) and 16.58 % (average 14.85 %), respectively, while for 1.5 ratio the of metallic Ni influence delivers lower PCE devices 14.90 % (average 13.50 %).

## 4. Discussion

In this work, we examined the solution combustion synthesis of Cu:NiO$_x$ films by using different molar ratios (w/o, 0.1 and 1.5) of fuel acetylacetone (Acac) to oxidizer (nitrates) precursors as well as various thermal processing annealing temperatures (150, 200 and 300 °C). Thermogravimetric analysis (TGA and DTA) results showed that complete combustion process at ~150 °C can occur in bulk analogues. XRD measurements revealed that the corresponding Cu:NiO$_x$ films crystallize to NiO phase upon annealing temperature at 300 °C, irrespectively to Acac concentration, while for lower annealing temperatures (150, 200 °C) no crystal phase was observed. XPS, AFM, UV-Vis spectroscopy and contact angle measurements on the films strongly support the incomplete combustion of the Cu:NiO$_x$ thin films for annealing temperatures at 150 °C and 200 °C. XPS measurements on Cu:NiO$_x$ film revealed the presence of high atomic ratio of remnants for thermal annealing at 200 °C which are remarkably reduced for films annealed at 300 °C. Surface topography images and thickness measurements via AFM and profilometer showed that the Cu:NiO$_x$ films annealed at 300 °C have a lower thickness (~50 nm) and roughness (~0.7 nm) compared to ~80 nm thickness and ~1.5 nm roughness for the Cu:NiO$_x$ films annealed at temperatures 150 °C due to remnants in the film. Moreover, Cu:NiO$_x$ films annealed at 300 °C have an improved hydrophilicity, showing a contact angle of 20 deg., while the films annealed at 150 °C and 200 °C show angles more than 60 deg due to surface remnants. Regarding the optical absorption measurements, the 300 °C thermally annealed films exhibit a distinct absorption curve ascribed to the formed crystalline Cu:NiO$_x$, while the lower annealing temperature films at 150 °C and 200 °C lack any strong absorption in the range of measured wavelengths due to amorphous phase; these results are consistent with the paper reported XRD findings. To conclude, the presented solution combustion chemistry findings in Cu:NiO$_x$ thin films are confirmed by applying the various ratios of Acac/Oxidizer and annealing processing temperatures of SCS based Cu:NiO$_x$ HTLs in triple cation based Cs$_{0.04}$(MA$_{0.17}$FA$_{0.83}$)$_{0.96}$Pb(I$_{0.83}$Br$_{0.17}$)$_3$ inverted PVSCs. The Cu:NiOx HTLs annealed at temperatures 150 °C and 200 °C irrespective to Acac/Oxidizer ratios provided limited functionality PVSCs due to incomplete combustion process that resulted to amorphous Cu:NiO$_x$ with remnants as confirmed by the presented XRD and XPS measurements respectively. The crystalline phase of Cu:NiO$_x$ HTLs and efficient inverted PVSCs performance obtained at annealing temperature of 300 °C irrespective to Acac/Oxidizer ratio. Following the solution combustion synthesis route that have been investigated within this manuscript, the Cu:NiO$_x$ crystalline HTLs annealed at 300 °C with 0.1 ratio of Acac/oxidizer resulted to 16.58 % PCE for the triple cation based Cs$_{0.04}$(MA$_{0.17}$FA$_{0.83}$)$_{0.96}$Pb(I$_{0.83}$Br$_{0.17}$)$_3$ inverted PVSCs.

**Funding:** This research was funded by the European Research Council (ERC) under the European Union's Horizon 2020 research and innovation program, (H2020 European Research Council, grant number 647311), and further supported from the academic yearly research activity internal Cyprus University of Technology budget.

# Supporting information

# Thermal Analysis of Metal-Organic Precursors for Functional Cu:NiOx Hole Transporting Layer in Inverted Perovskite Solar Cells: Role of Solution Combustion Chemistry in Cu:NiOx Thin Films Processing

*Apostolos Ioakeimidis,[a] Ioannis T. Papadas,[a,c] Eirini D. Koutsouroubi,[b] Gerasimos S. Armatas[b] and Stelios A. Choulis[a]\**

[a] Molecular Electronics and Photonics Research Unit, Department of Mechanical Engineering and Materials Science and Engineering, Cyprus University of Technology, Limassol, Cyprus.

[b] Department of Materials Science and Technology, University of Crete, Heraklion 70013, Greece.

[c] Department of Public and Community Health, School of Public Health, University of West Attica, Athens, Greece.

*Corresponding Author: Prof. Stelios A. Choulis, E-mail: stelios.choulis@cut.ac.cy

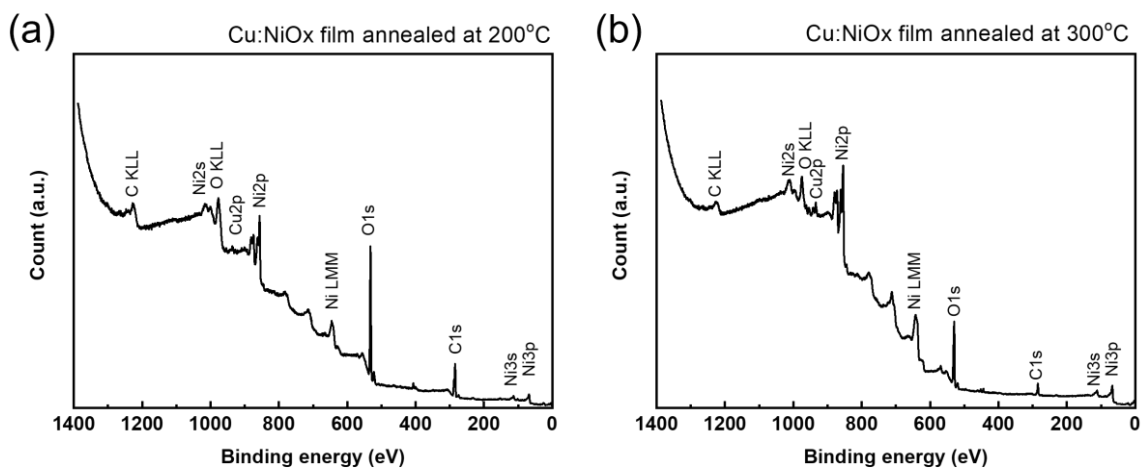

**Figure S1.** XPS survey spectra of the Cu:NiOx films fabricated form precursor containing 0.1 Acac and annealed at (a) 200 °C and (b) 300 °C.

**Table S1.** XPS calculated atomic ratios for the Cu:NiOx films fabricated form precursor containing 0.1 Acac and annealed at 200 and 300 °C.

| Sample | C[a] (%) | N[b] (%) | Ni (%) | Cu (%) | O (%) |
|---|---|---|---|---|---|
| 200 °C annealed Cu:NiOx film | 18.40 | 8.05 | 17.59 | 2.10 | 53.86 |
| 300 °C annealed Cu:NiOx film | 3.26 | 1.45 | 42.11 | 5.22 | 47.95 |

[a]C-oxygen/nitrogen bonded species corresponding to a 6.51 and 0.64 wt.% content for 200 and 300 °C annealed Cu:NiOx films, respectively. [b]Nitrogen-containing species corresponding to a 3.32 and 0.33 wt.% content for 200 and 300 °C annealed Cu:NiOx films, respectively.

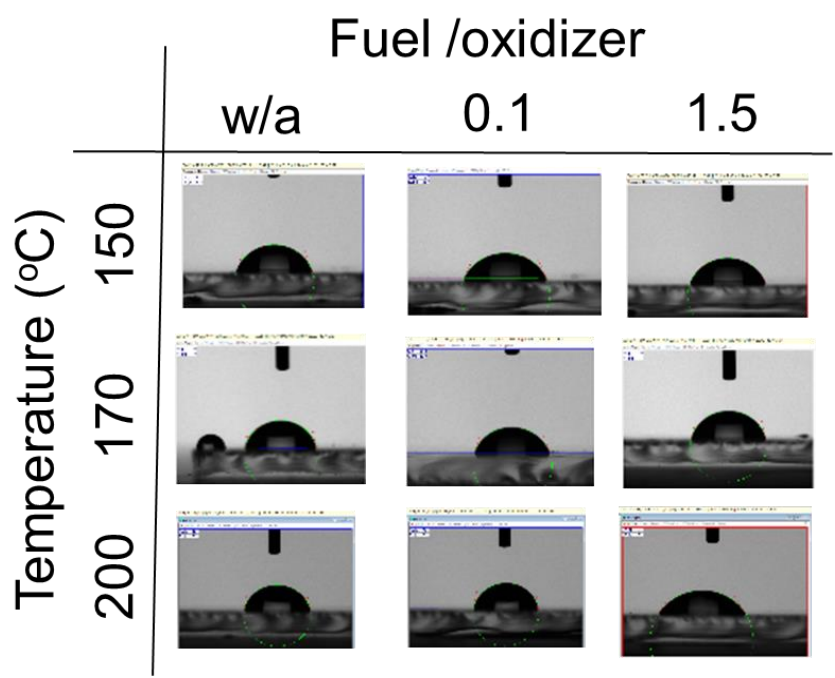

**Figure S2.** contact angle picture of water on films prepared using precursor containing w/o, 0.1 and 1.5 molar ratio of fuel (Acac) to oxidizer (Cu, Ni nitrates) annealed at 150, 170 and 200 °C, respectively.

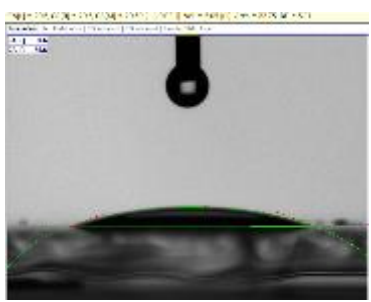

**Figure S3.** contact angle picture of water on films prepared using precursor 0.1 molar ratio of fuel (Acac) to oxidizer (Cu, Ni nitrates) annealed at 300 °C showing an angle of 20 deg.